\documentclass{birkjour}
 \newtheorem{theorem}{Theorem}[section]
 \newtheorem{cor}[theorem]{Corollary}

 \theoremstyle{definition}
 
 \theoremstyle{remark}

 \newtheorem{problem}{Problem}
 \numberwithin{equation}{section}

\usepackage{hyperref}

\usepackage{amsmath,amsthm}
\usepackage{amssymb}

\usepackage{enumitem}

\usepackage{graphicx}

\usepackage{caption}
\usepackage{subcaption}

\usepackage[T1]{fontenc}
\usepackage{amsfonts}
\usepackage{float}
\usepackage{cancel}
\begin{document}

\title[The dual of Philo's shortest line segment problem]
{The dual of Philo's shortest line segment\\ problem}

%----------Author 1
\author{Yagub N. Aliyev}

\address{%
School of IT and Engineering\\ 
ADA University\\
Ahmadbey Aghaoglu str. 61 \\
Baku 1008, Azerbaijan}

\email{yaliyev@ada.edu.az}

\thanks{This work was supported by ADA University Faculty Research and Development Fund.}



\date{January 1, 2004}
%\dedicatory{}
%%% ----------------------------------------------------------------------

\begin{abstract}
We study the dual of Philo's shortest line segment problem and find the optimal line segments passing through two given points, with a common endpoint, and with the other endpoints on a given line. This problem is dual, in a point-and-line-exchanging sense, to a famous problem of antiquity used to solve the problem of duplicating the cube. The provided solution uses multivariable calculus and elementary geometry methods. Interesting connections with the angle bisector of the triangle are explored. A generalization of the problem using $l_p$ ($p\ge 1$) norm is proposed. The particular case $p=\infty$ is also studied. It is shown that in the cases $p=0$ and $p=2$ the median and the symedian, respectively, of a triangle do not always give a solution for the corresponding optimization problems. The general case $p\ne 1$ and related problems are proposed as open problems.
\end{abstract}

%%% ----------------------------------------------------------------------
\maketitle
%%% ----------------------------------------------------------------------
%\tableofcontents
\section{Introduction} 
\begin{figure}[htbp]
\centerline{\includegraphics[scale=.15]{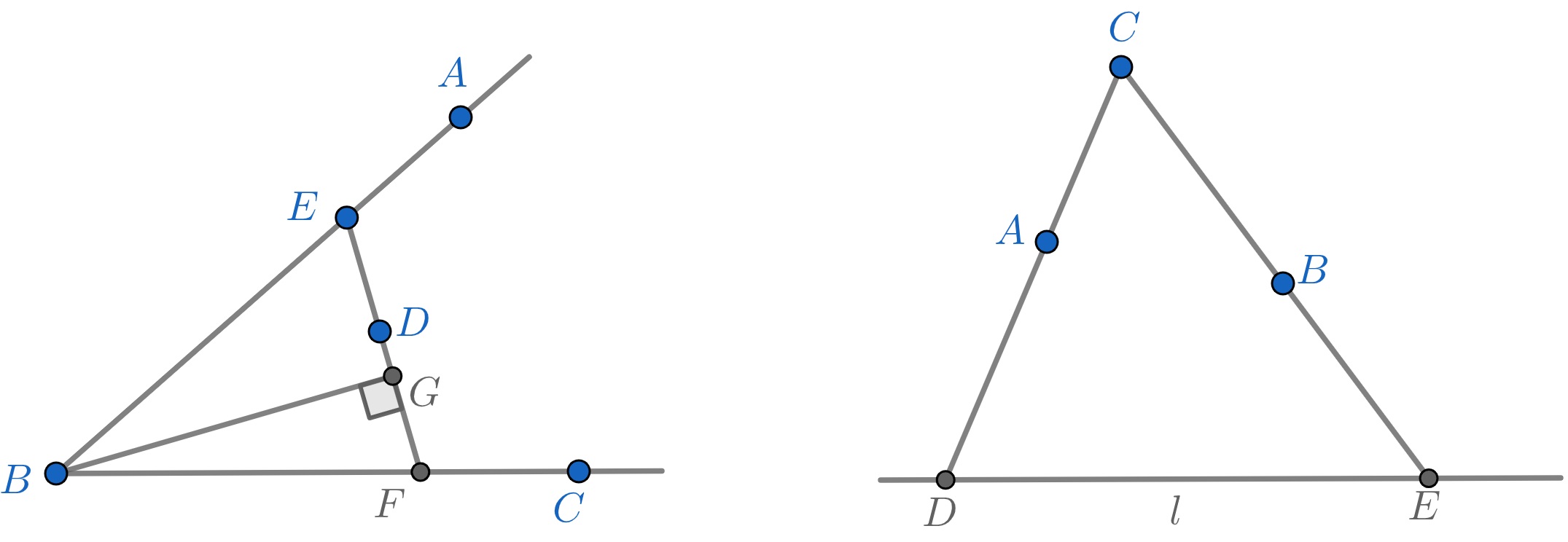}}
\caption{Philo's line and the dual problem}
\label{fig0}
\end{figure}

Maximum-minimum problems in geometry played crucial role in the history of development of mathematical methods \cite{courant}, Chapter VII, \cite{bolt}, \cite{polya} Chapters VIII-XIX. Many attempts to solve variety of geometry problems for optimization lead to the develoment of the differential and integral calculus and later to the calculus of variations. Geometric extrema problems play a crucial role in mathematics education and research \cite{andre}, \cite{tikho}, Chs. 4 and 13; \cite{zet}; \cite{mit}; \cite{cox1}, Sect 1.8. In the mathematical olympiads for schools or universities, and the journals with problem solving columns such as Mathematics Magazine, The American Mathematical Monthly, and Crux Mathematicorum one can often find at least one geometric inequality included in its list of proposed problems. The recent breakthough in automation of proofs for classical euclidean geometry theorems is yet to be done for the geometric inequalities \cite{trinh}. The connections of the geometric extrema problems with $l_p$ norm (see for example \cite{lust}, p. 20) in functional analysis were discussed in \cite{prot1}, \cite{zas}. Historically, mathematicians recognized that certain geometric optimization problems lack solutions and even when they do have a solution it is not always possible to construct it with the euclidean instruments (i.e. a pair of compasses and a straightedge). The problem about the minimal line segment with a fixed point and inscribed into a given angle is one of them \cite{booth}, p. 405-406, \cite{prot}, p. 36-37, \cite{shklyar} p. 33, 179-181, \cite{neov}, \cite{anghel2}, \cite{neub}. In this problem an angle $ABC$ and a point $D$ in it is given and it is asked to find the line $EF$ through this point which has shortest line segment $EF$ cut by this angle (see Appendix). Sir I. Newton mentioned this problem and its generalizations as an example for his method of fluxions (see \cite{newton}, p. 45, Problem 5.III). The solution of this problem is known in the literature as Philo's line \cite{eves}, \cite{eves1} p. 39 and 234-238, \cite{wells} p. 182-183, \cite{weis}, \cite{kimb} p. 115–116. If $BG$ is the perpendicular to $EF$ from point $B$, then $EF$ is shortest when $|ED|=|FG|$ (see Figure \ref{fig0}).

Philo or Philon of Byzantium found the solution of the classical problem of the duplication of a cube using this line \cite{bollo}, p. 107, \cite{wiki}, \cite{pras} p. 24, \cite{tous}. The digits of the length of Philo's line through the incenter of triangle with sidelengths 3,4,5 appear in the OEIS as A195284 \cite{kimb1}. Generalizations for the non-euclidean geometry and the higher dimensions appeared in \cite{cox}, \cite{wetter}. Related problem of the longest or shortest chords of conics and curves in general have similar solutions \cite{anghel1}, \cite{anghel3}, \cite{aliyev}.

There are examples of the use of the principle of duality in metric geometry \cite{mord} but the duality is usually used in projective geometry to explain the similarity between the theorems where the roles of lines and points are interchanged. If we use the analogous idea in the above extremum problem, then we obtain the following dual problem in a point-and-line-exchanging sense (\cite{aliyev}, p. 4). Given the line $l$ and the points $A$ and $B$ find the point $C$ with the rays $CA$ and $CB$ intersecting the line $l$ at the points $D$ and $E$, respectively, such that the sum of the lengths of the line segments $CD$ and $CE$ is minimal (see Figure \ref{fig0} and Table 1). Note that  the line segments $AC$ and $BC$ themselves do not intersect the line $l$ or equivalently $C$ is on the same open half plane as $A$ and $B$. Furthermore, the points $A$ and $B$ are located between $l$ and $C$. The problem presents a significantly greater challenge compared to Philo’s shortest line problem. The problem is solved using a combination of multivariable calculus and elementary geometric methods. The paper concludes with noteworthy formulas related to the configuration and some open problems for further studies.

\noindent
\begin{table}

\begin{tabular}{ |c||c|c|}

 \hline
 & Philo's line problem & Dual problem\\
 \hline
Fixed point(s) & 1 ($D$) & 2 ($A$ and $B$) \\
Fixed line(s) & 2 ($BA$ and $BC$) & 1 ($l$)\\
Moving point(s) & 2 ($E$ and $F$) & 1 ($C$)\\
Moving line(s) & 1 ($EF$) & 2 ($CD$ and $CE$)\\
\hline

\end{tabular}
\label{tab1}
\caption{The number of elements in Philo's line problem and its dual.}
\end{table}
\section{Main results.}

As was mentioned before the optimal solution of the geometric extrema problems can sometimes be non-constructible with the help of a pair of compasses, a ruler, and the given data. Assuming existence of optimal configuration without a proper proof can sometimes lead to contradictory results (see \cite{radem}, Sect. 18a in the original and Sect. 21 in later editions). See also for example \cite{lust}, p. 56-57 where it is shown how assuming existence of a triangle with maximal sum of angles in non-euclidean geometry leads to a "proof" of the fifth postulate of Euclid. Therefore, without a justification of the existence of the optimal solution the proof is incomplete (cf. \cite{niven}, Chs. 4 and 12). For example, in Hadamard's classical geometry problem book some of such extrema problems are provided with a comment like "assume that a maximum exists" (see e.g. Exercise 366 in p. 313, Exercise 418b in p. 322 of \cite{hadam}). Most of the time it is possible to refer to calculus results for more rigorous proof with compact sets and Weierstrass theorem but it makes the solution less appealing from the geometric point of view.  Another possible way out from this situation is to give or construct the optimal solution first in the diagram and then ask to prove that the other possibilities are not optimal. In \cite{fejes}, Sect. 1.3 these two approaches to geometric extrema problems are referred to as \textit{indirect} and \textit{direct} proofs, respectively (cf. \cite{shklyar}, pp. 7-10). In the problem about Philo's line the direct proof would be giving $\triangle BEF$ and points $D$ and $G$ on the side $EF$ such that $|ED|=|FG|$, $BG\perp EF$, and then proving that any other line segment $E_1F_1$ through point $D$ has a greater length (see Appendix). We can formulate the mentioned dual problem above in the same way by giving the optimal solution first. The proof establishes the optimality of the solution and its derivation.

\begin{figure}[htbp]
\centerline{\includegraphics[scale=.15]{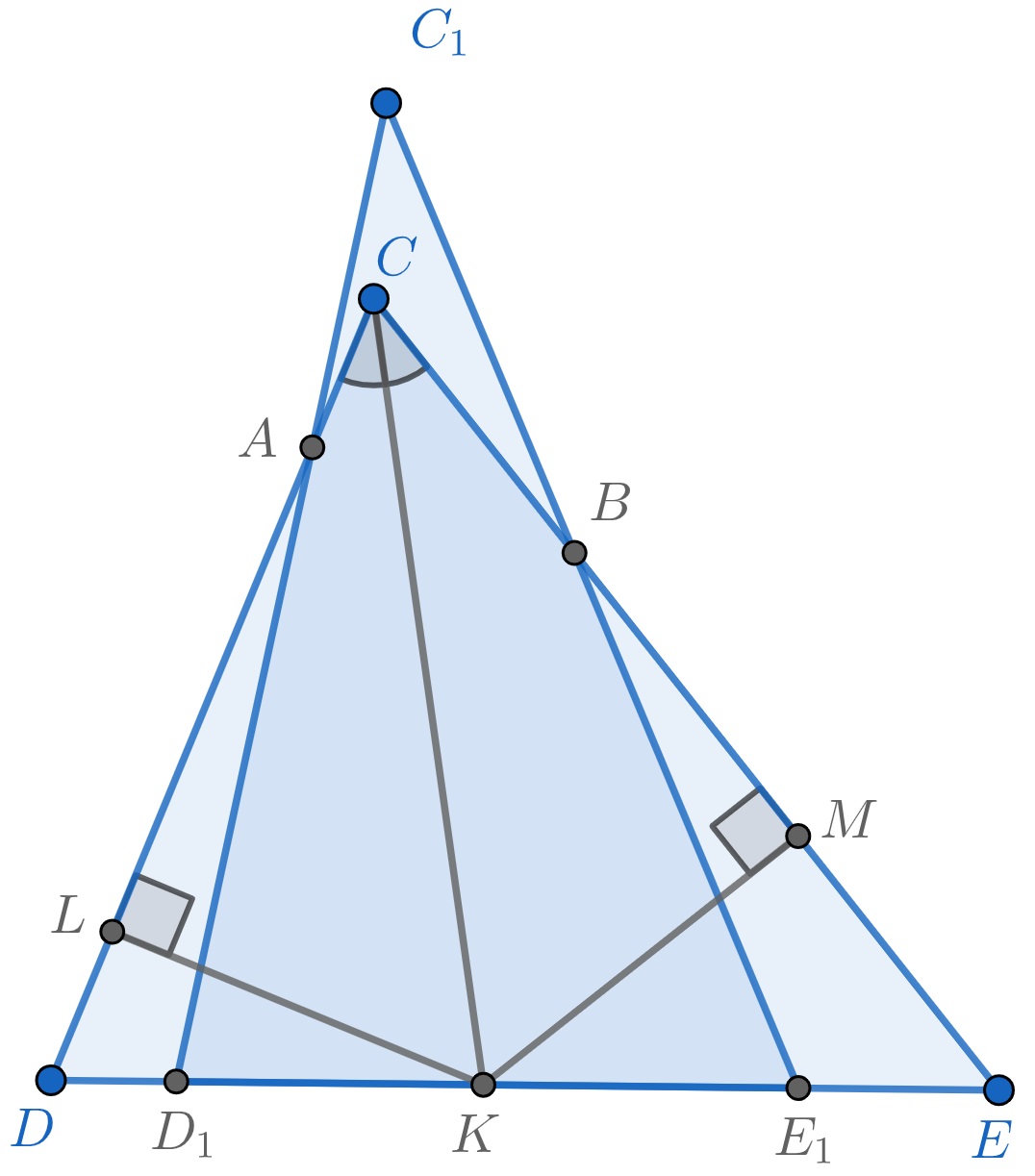}}
\caption{Inequality  $|C_1 D_1 |+|C_1 E_1 |>|CD|+|CE|$}
\label{fig1}
\end{figure}

\begin{theorem} Let $CDE$ be a triangle with acute angles at the vertices $D$ and $E$. Let $CK$ be the angle bisector of the triangle $CDE$. Drop the perpendiculars $KL$ and $KM$ to the sides $CD$ and $CE$, respectively. Take the points $A$ and $B$ on the sides $CD$ and $CE$, respectively, such that $|AC|=|LD|$, $|BC|=|ME|$. Then for any point $C_1$ different from $C$, such that the rays $C_1 A$ and $C_1 B$ intersect the line $DE$ at the points $D_1$ and $E_1$, respectively, inequality $|C_1 D_1 |+|C_1 E_1 |>|CD|+|CE|$ holds true (see Figure \ref{fig1}).
\end{theorem}

\begin{proof} Let us denote the coordinates of the points as
\(A(a,b),\ B(c,d),\ D_{1}(x_{1},0)\) and \(E_{1}(x_{2},0)\). Without loss of generality we can assume that $a< c$ and $b\ge d>0$. If we
denote coordinates of \(C_{1}\ \)as \((x,y)\), then \(x_{1} = \frac{ay - bx}{y - b}\) and
\(x_{2} = \frac{cy - dx}{y - d}\). Here \((x,y)\in \mathbb{D}\), where
\(\mathbb{D}=\{(x,y)| x\in( - \infty, + \infty),\ y>b\}\). The last inequality is needed to guarantee the condition in the statement of the problem which says that the rays $C_1 A$ and $C_1 B$ intersect the line $DE$. We obtain that
$$
\left| C_{1}D_{1} \right| + \ \left| C_{1}E_{1} \right| = \sqrt{\left( x - x_{1} \right)^{2} + y^{2}} + \sqrt{\left( x - x_{2} \right)^{2} + y^{2}}
$$
$$
= y\left( \sqrt{\left( \frac{x - a}{y - b} \right)^{2} + 1} + \sqrt{\left( \frac{x - c}{y - d} \right)^{2} + 1} \right) =: f(x,y).
$$
Note that the function \(f(x,y)\) is continuous in $\mathbb{D}$ and \(f(x,y)\rightarrow + \infty\) as
\(x \rightarrow \pm \infty\),
\(y \rightarrow  b\) when $x\ne a$. Similarly, \(f(x,y)\rightarrow + \infty\) as \(y \rightarrow +\infty\). Also \(\lim_{(x,y)\rightarrow (a,b)}f(x,y) \) does not exist and depends on the direction of approach to the point \((a,b)\). Indeed, if $b>d$, then
$$
\lim_{t\rightarrow 0^+}f(a+k_1t , b+k_2t) =b \left(\sqrt{\frac{k_1^{2}+k_2^{2}}{k_2^{2}}}+\frac{\sqrt{(b-d)^{2}+\left(a-c\right)^{2}}}{b-d}\right),
$$
where $k_2>0$. This limit is minimal when $k_1=0$ which corresponds to the direction of the line $x=a$. If $b=d$, then the last limit is also infinite. Function \(f(x,y)\) is also bounded from
below because it is a positive function. Then either there is a point \((x^{*},y^{*})\in \mathbb{D}\) where function \(f(x,y)\) is minimal or this function approaches its infimum when \((x,y)\rightarrow (a,b)\) \((y>b)\). We will study the last case later. In the first case \(f_x(x^{*},y^{*}) =0\), \(f_y(x^{*},y^{*}) = 0\). It is not hard to compute that $f_x$ is negaitve if $x < a$ and positive if $x > c$. Therefore, if \(x<a\), then \(f(x,y)>f(a,y)\), and if \(x>c\), then \(f(x,y)>f(c,y)\), and consequently \(a\le x^{*}\le c\). Note also that if $y$ is fixed, then the function $f(x,y)$ is convex with respect to $x$. Indeed,
$$
f_{xx}=\frac{y(y-b)} {\sqrt{{\left(\left(x-a\right)^{2}+(y-b)^{2}\right)^{3}} }}+\frac{y\left(y-d\right)} {\sqrt{{\left(\left(x-c\right)^{2}+(y-d)^{2}\right)^{3}} }}>0.
$$
We calculate 
$$
f_{x}(a,y)=\frac{ y\left(a-c\right)}{\left(y-d\right)\sqrt{{\left(a-c\right)^{2}+\left(y-d\right)^{2}}}}<0,
$$
$$
f_{x}(c,y)=\frac{ y\left(c-a\right)}{\left(y-b\right)\sqrt{{\left(a-c\right)^{2}+\left(y-b\right)^{2}}}}>0.
$$
Therefore \(a<x^{*}<c\). In order to locate the point $x^{*}$, note that equality \(f_x(x,y) =0\) implies
$$\frac{x - a}{(y - b)^{2}\sqrt{\left( \frac{x - a}{y - b} \right)^{2} + 1}} = \frac{c - x}{(y - d)^{2}\sqrt{\left( \frac{x - c}{y - d} \right)^{2} + 1}}
\Leftrightarrow 
$$
$$\frac{x - a}{(y - b)\sqrt{(x - a)^{2} + (y - b)^{2}}} = \frac{c - x}{(y - d)\sqrt{(x - c)^{2} + (y - d)^{2}}}.$$
\begin{figure}[htbp]
\centerline{\includegraphics[scale=.15]{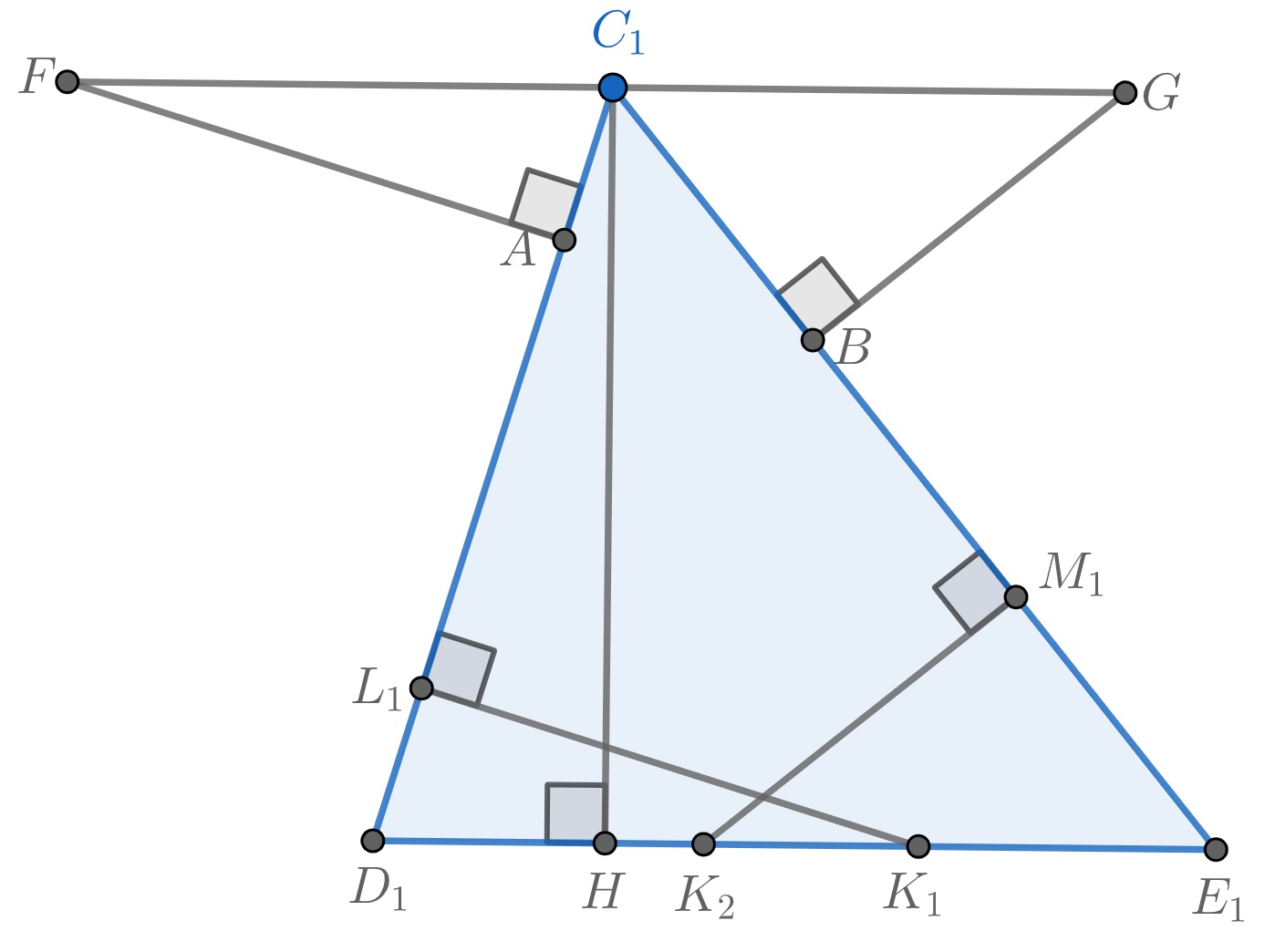}}
\caption{$|K_1L_1|=|K_2M_1|.$}
\label{fig2}
\end{figure}
The last equality can be interpreted geometrically in the following way. Take the points $L_1$ and $M_1$ on the sides $C_1D_1$ and $C_1E_1$, respectively, such that $|AC_1|=|L_1D_1|$, $|BC_1|=|M_1E_1|$. Take the points $K_1$ and $K_2$ on the line $D_1E_1$ such that $K_1L_1\perp C_1D_1$ and $K_2M_1\perp C_1E_1$. The equality \(f'_x(x,y) =0\) is equivalent to say that $|K_1L_1|=|K_2M_1|$ (see Figure \ref{fig2}). Indeed, if we draw the line through $C_1$ and parallel to $D_1E_1$ and take the points $F$ and $G$ such that $AF\perp AC_1$ and $BG\perp BC_1$, then $|AF|=|K_1L_1|$ and $|BG|=|K_2M_1|$. On the other hand,
$$
|AF|=|AC_1|\cdot \tan{\angle AC_1F}=\sqrt{(x - a)^{2} + (y - b)^{2}}\cdot \frac{y-b}{a-x}.
$$
Similarly,
$$
|BG|=\sqrt{(x - c)^{2} + (y - d)^{2}}\cdot \frac{y-d}{c-x}.
$$
Therefore (cf. \cite{aliyev1}, Problem 3) $$|K_1L_1|=|K_2M_1|.\eqno(1)$$

The other condition (\(f_y(x,y) =0\)) can also be interpreted geometrically but its little harder than the previous condition. First, drop the perpendicular $C_1H$ to the side $D_1E_1$. Next, we calculate
$$
f_y(x,y) =\left( \sqrt{\left( \frac{x - a}{y - b} \right)^{2} + 1} + \sqrt{\left( \frac{x - c}{y - d} \right)^{2} + 1} \right) +
$$
$$y\left( -\frac{\left(x-a\right)^{2}}{\sqrt{1+\frac{\left(x-a\right)^{2}}{\left(y-b\right)^{2}}}\, \left(y-b\right)^{3}}-\frac{\left(x-c\right)^{2}}{\sqrt{1+\frac{\left(x-c\right)^{2}}{\left(y-d\right)^{2}}}\, \left(y-d\right)^{3}} \right).
$$
If \(a< x< c\), then the equality $
f_y(x,y) =0$ can be interpreted as
$$
|C_1D_1|+|C_1E_1|=\frac{|D_1H|^2}{|L_1D_1|}+\frac{|E_1H|^2}{|M_1E_1|}.\eqno(2)
$$
Indeed, first note that 
$$
\frac{1}{y}\left( |C_1D_1|+|C_1E_1| \right)=\sqrt{\left( \frac{x - a}{y - b} \right)^{2} + 1} + \sqrt{\left( \frac{x - c}{y - d} \right)^{2} + 1}.
$$
On the other hand note that
$$
\frac{\left(x-a\right)^{2}}{\sqrt{1+\frac{\left(x-a\right)^{2}}{\left(y-b\right)^{2}}}\, \left(y-b\right)^{3}}+\frac{\left(x-c\right)^{2}}{\sqrt{1+\frac{\left(x-c\right)^{2}}{\left(y-d\right)^{2}}}\, \left(y-d\right)^{3}}
$$
$$
=\frac{\frac{\left(x-a\right)^{2}}{\left(y-b\right)^{2}}}{\sqrt{{\left(x-a\right)^{2}}+{\left(y-b\right)^{2}}}}+\frac{\frac{\left(x-c\right)^{2}}{\left(y-d\right)^{2}}}{\sqrt{{\left(x-c\right)^{2}}+{\left(y-d\right)^{2}}}}
$$
$$
=\frac{\tan^2{\angle AC_1F}}{|AC_1|}+\frac{\tan^2{\angle BC_1G}}{|BC_1|}.
$$
We know that $\angle AC_1F=\angle D_1$, $\angle BC_1G=\angle E_1$, $|AC_1|=|L_1D_1|$ and $|BC_1|=|M_1E_1|$. Therefore,
$$
|C_1D_1|+|C_1E_1|=|C_1H|^2\left(\frac{\tan^2{\angle D_1}}{|L_1D_1|}+\frac{\tan^2{\angle E_1}}{|M_1E_1|}\right),
$$
\begin{figure}[htbp]
\centerline{\includegraphics[scale=.15]{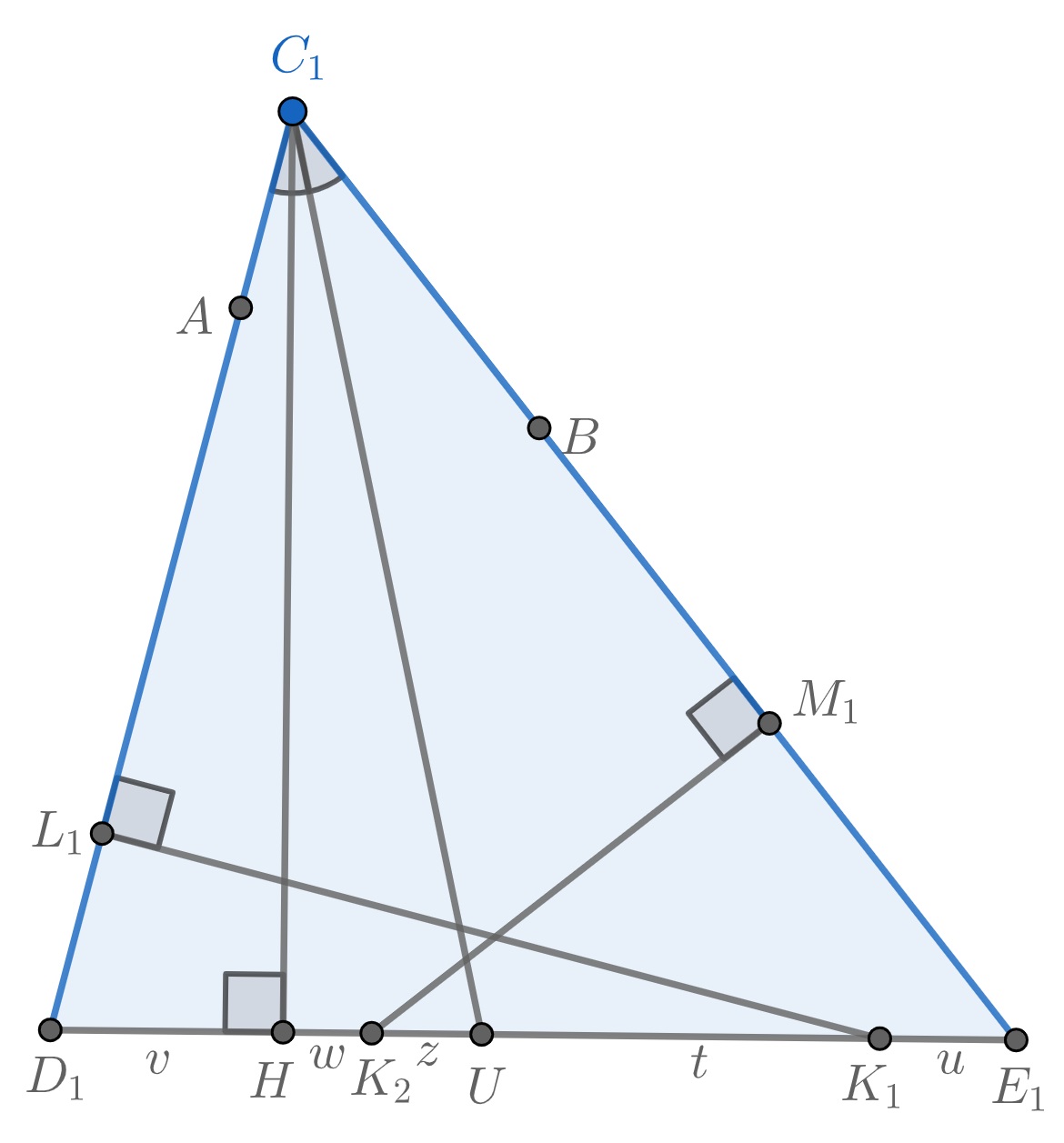}}
\caption{One possible order of points on line $D_1E_1$.}
\label{fig3}
\end{figure}
from which it follows that (2) holds true. Since $\frac{|D_1H|^2}{|L_1D_1|}=|D_1H|\cdot \frac{|C_1D_1|}{|D_1K_1|}$ and $\frac{|E_1H|^2}{|M_1E_1|}=|E_1H|\cdot\frac{|C_1E_1|}{|E_1K_2|}$, the equality (2) can also be written as
$$
|C_1D_1|+|C_1E_1|=|C_1D_1|\cdot \frac{|D_1H|}{|D_1K_1|}+|C_1E_1|\cdot\frac{|E_1H|}{|E_1K_2|},
$$
which we can rewrite as
$$
|C_1D_1|\cdot \left(1- \frac{|D_1H|}{|D_1K_1|}\right)=|C_1E_1|\cdot\left(\frac{|E_1H|}{|E_1K_2|}-1\right), \eqno(3)
$$
or finally as
$$
|C_1D_1|\cdot \frac{|K_1H|}{|D_1K_1|}=|C_1E_1|\cdot\frac{|K_2H|}{|E_1K_2|}.
$$
From this we obtain
$$
\frac{|D_1C_1|}{|C_1E_1|}\cdot \frac{|E_1K_2|}{|K_2H|}\cdot\frac{|HK_1|}{|K_1D_1|}=1. \eqno(4)
$$

If $AU$ is the angle bisector of the triangle $D_1C_1E_1$, then by the angle bisector theorem $\frac{|D_1C_1|}{|C_1E_1|}=\frac{|D_1U|}{|UE_1|}$. By noting this we can write (4) as
$$
\frac{|D_1U|}{|UE_1|}\cdot \frac{|E_1K_2|}{|K_2H|}\cdot\frac{|HK_1|}{|K_1D_1|}=1. \eqno(5)
$$
In the following the
proof will break into three cases (Figures \ref{fig3}, \ref{fig4}, and \ref{fig5}) depending on the relative positions of
$K_1$ and $K_2$. If we denote $|D_1H|=v$, $|K_2H|=w$, $|K_2U|=z$, $|K_1U|=t$, $|K_1E_1|=u$ (see Figure \ref{fig3}), equation (5) can be written simply as
$$
\frac{v+w+z}{t+u}\cdot \frac{u+t+z}{w}\cdot\frac{w+z+t}{v+w+z+t}=1,
$$
which after simplifications gives
$$
t^{2} v+t^{2} z+t u v+t u z+2 t v z+2 t w z+2 t z^{2}
$$
$$
+u v z+u w z+u z^{2}+v w z+v z^{2}+w^{2} z+2 w z^{2}+z^{3}
$$
$$
=t(t v+t z+u v+u z+2 v z+2 w z+2 z^{2})
$$
$$
+z(u v+u w+uz+v w+v z+w^{2} +2 w z+z^{2})=0.
$$
It is possible only when $z=t=0$, which means that the points $K_1,\ K_2,$ and $U$ coincide. The other cases when the points $K_1$ or $K_2$ are outside of the side $D_1E_1$, or when the point $K_1$ is closer to the point $D_1$ than the point $K_2$ can be considered similarly. For example, if the order of the points is as in Figure \ref{fig4}, where the notations for the lengths of the line segments has slightly changed, then
$$
\frac{v+w+t}{z+u}\cdot \frac{u}{z+t+w}\cdot\frac{w}{v+w}=1,
$$
which is not possible. Indeed, in this case since $\frac{v+w+t}{z+u}=\frac{|D_1C_1|}{|C_1E_1|}<1,$  $v+w+t<z+u$, and since $$u=|K_2M_1|\cdot\csc{\angle E_1}=|K_1L_1|\cdot\csc{\angle E_1}<|K_1L_1|\cdot\csc{\angle D_1}=v+w,$$  $u<v+w$, and therefore 
$$
\frac{v+w+t}{z+u}\cdot \frac{u}{v+w}\cdot\frac{w}{z+t+w}<1.
$$
\begin{figure}[htbp]
\centerline{\includegraphics[scale=.15]{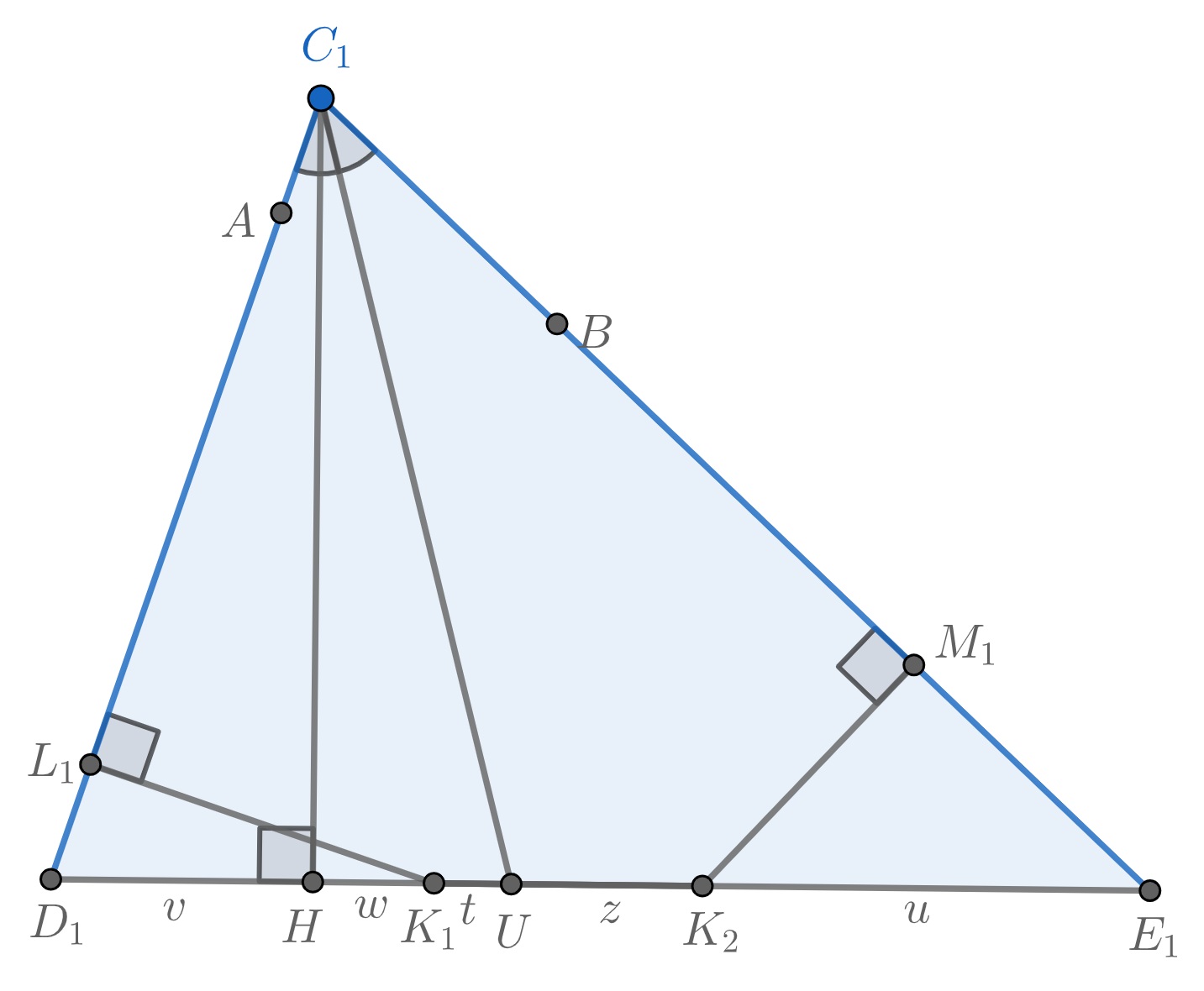}}
\caption{Another order of points on line $D_1E_1$.}
\label{fig4}
\end{figure}
Finally, if the order of the points is as in Figure \ref{fig5}, where again the notations for the lengths of the line segments has slightly changed, then
$$
\frac{w+v+z}{t}\cdot \frac{t+z}{v}\cdot\frac{v+z+t+u}{w+v+z+t+u}=1,
$$
which again simplifies to an impossible equality
$$
t^{2} w+t^{2} z+t u w+t u z+2 t v z+2 t w z+2 tz^{2}
$$
$$
+u v z+u w z+uz^{2}+v^{2} z+v w z+2 vz^{2}+wz^{2}+z^{3}=0.
$$
\begin{figure}[htbp]
\centerline{\includegraphics[scale=.15]{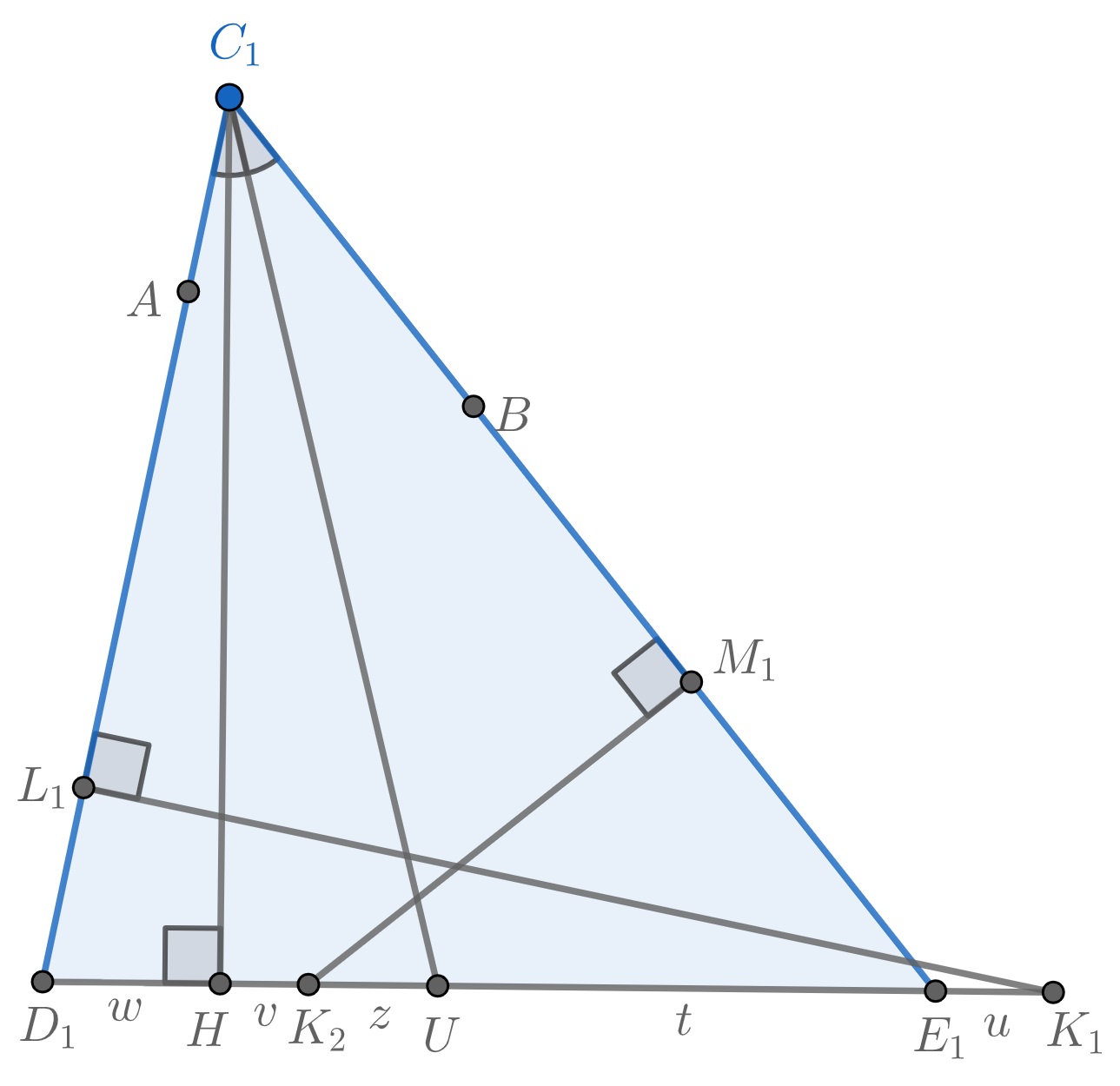}}
\caption{One more order of points on line $D_1E_1$.}
\label{fig5}
\end{figure}
Note that because of the equality (1), we only need to consider the cases when the point $U$ is between the points $K_1$ and $K_2$. Otherwise, if $U$ is not between the points $K_1$ and $K_2$, then one can drop perpendiculars from the point $U$ to the sides $C_1D_1$ and $C_1E_1$ and one of them will be larger than the other one, which is impossible. Similarly, we do not need to consider the cases when the points $K_1$ and $K_2$ are on the different sides of point $H$, because in these cases one of the sides of the equality (3) is positive while the other side is negative and therefore (3) does not hold.

If $\angle D_1=\frac{\pi}{2}$ ($x=a$), then the point $K_1$ goes to infinity and the equation (5) simplifies to 
$
\frac{|D_1U|}{|UE_1|}\cdot \frac{|E_1K_2|}{|K_2H|}=1,
$
which simply means that the points $K_2$ and $U$ coincide. Similarly, if $\angle E_1=\frac{\pi}{2}$ ($x=c$), then the point $K_2$ goes to infinity and the equation (5) simplifies to 
$
\frac{|D_1U|}{|UE_1|}\cdot\frac{|HK_1|}{|K_1D_1|}=1,
$
which implies that the points $K_1$ and $U$ coincide.

We proved that if \(C_1(x^{*},y^{*})\) is a point with\(f_x=f_y=0\), then the endpoint $U$ of the angle bisector $C_1U$ coincides with points $K_1,K_2$. Hence, the optimal point will either be such a point, if it exists, or the point $C_1$ where the
deriviatives are undefined. The second case is not possible because $f_x$ and $f_y$ are defined for all points of $\mathbb{D}$. Note that triangle $CDE$ and its angle bisector $CK$ satisfy all these conditions together with (1) and (2). In order to complete the proof, we need to show that there are no minimum points other than $C$.

First, note that if $CK$ is the
angle bisector, and $KL$ and $KM$ are the perpendiculars to the sides (see Fig. 2),
then $CL$ and $CM$ are congruent. Hence, if $AC=LD$ and $BC=ME$, then $AD=BE$.
We will use this fact to study the locus of the intersection point of lines $L_1K_1$ and $M_1K_2$ when $|AD_1|=|BE_1|$ and $D_1(x_1,0)$ changes on line $DE$, and then intersection of this locus with line $DE$ (see Figure \ref{fig6}). If $b=d$, then by the symmetry this locus is the perpedicular bisector of the line segment $AB$, which always intersect line $DE$ at a unique point. So, we can assume that $b>d$. Let us denote $|AD_1|=|BE_1|=t$. Then $x_1=a-\sqrt{t^2-b^2}$ and  $x_2=c+\sqrt{t^2-d^2}$. Then the equations of the lines $AD_1$ and $BE_1$ are $\frac{y-b}{x-a}=\frac{-b}{x_1-a}$ and $\frac{y-d}{x-c}=\frac{-d}{x_2-c}$, respectively. The coordinates of the intersection point $C_1(x_0,y_0)$ of these lines are
$$
{x_0}=-\frac{d\sqrt{t^{2}-b^{2}}\, \sqrt{t^{2}-d^{2}}+bd^{2}-bt^{2}}{t^{2} \left(b^{2}-d^{2}\right) \left(d^{2}-t^{2}\right)}\times
$$
$$
\times\left(\left(-c d\sqrt{t^{2}-d^{2}}-\left(b-d\right)\left(d^2-t^2\right) \right) \sqrt{t^{2}-b^{2}}+a b \left(d^2-t^2\right)\right),
$$
$$
{y_0}= \frac{ d b \left(\sqrt{t^{2}-d^{2}}\, b -d \sqrt{t^{2}-b^{2}}\right)\left(\sqrt{t^{2}-d^{2}}+\sqrt{t^{2}-b^{2}}-a +c \right)}{t^{2} \left(b^{2}-d^{2}\right)}.
$$
\begin{figure}[htbp]
\centerline{\includegraphics[scale=.15]{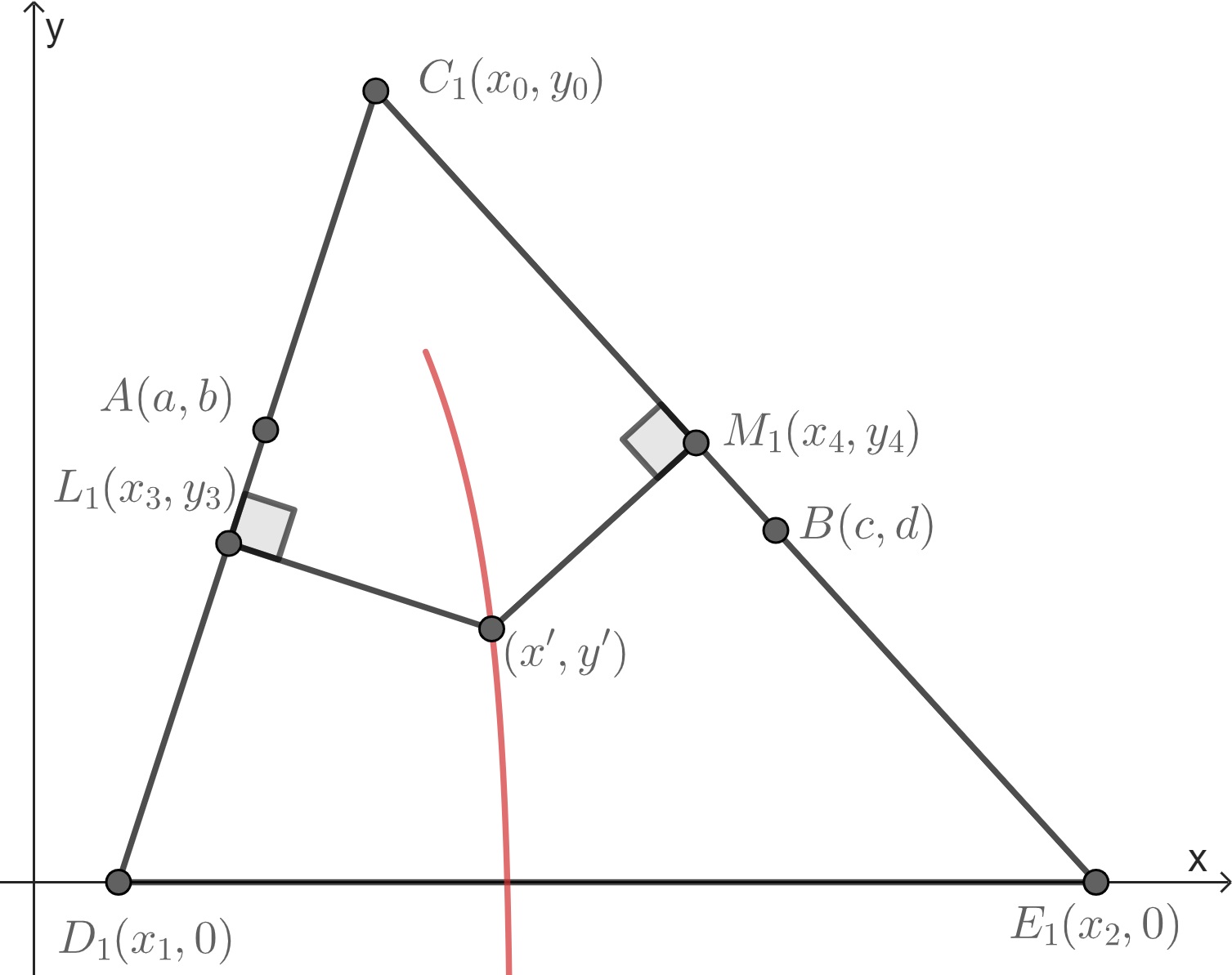}}
\caption{Locus of point $(x',y')$.}
\label{fig6}
\end{figure}
The coordinates of the point $L_1(x_3,y_3)$ are $x_3=x_0+x_1-a$, $y_3=y_0-b$. Similarly, the coordinates of the point $M_1(x_4,y_4)$ are $x_4=x_0+x_2-c$, $y_3=y_0-d$. Then the equations of the lines $L_1K_1$ and $M_1K_2$ are $\frac{y-y_3}{x-x_3}=\frac{x_1-a}{b}$ and $\frac{y-y_4}{x-x_4}=\frac{x_2-c}{d}$, respectively. By solving the last two equations for $y$ we calculate the ordinate $y'$ of intersection point of lines $L_1K_1$ and $M_1K_2$ as
$$
y'=\phi(t):=\frac{(bd-t^2)\left(\sqrt{1-\frac{b^2}{t^{2}}}+\sqrt{1-\frac{d^2}{t^{2}}}\right)+\frac{bd(c-a)}{t}}{b\sqrt{1-\frac{d^2}{t^{2}}}+d\sqrt{1-\frac{b^2}{t^{2}}}},
$$
where the denominator is always positive and the numerator is a decreasing function for $t\ge b$. It is because in the numerator of $\phi(t)$ the first term, which
is negative, is increasing in absolute value and the second term, which
is positive, is decreasing.

The abscissa $x'$ of this intersection point can also be determined but we will not need it:
$$
x'=-\frac{\left(b^{2} d-dt^{2}+b\sqrt{t^{2}-d^{2}}\, \sqrt{t^{2}-b^{2}}\right)}{\left(b^{2}-t^{2}\right) t^{2} \left(b^{2}-d^{2}\right)}\times
$$
$$
\times\left(\left(a b\sqrt{t^{2}-b^{2}}+\left(b^2-t^2\right) \left(b-d\right)\right) \sqrt{t^{2}-d^{2}}\\
+t^{2} \left(b-d\right) \sqrt{t^{2}-b^{2}}-c d \left(b^2-t^2\right)\right).
$$The formula for $x'$ and $y'$ determine a parametric curve which is the locus of intersection point of lines $L_1K_1$ and $M_1K_2$. This locus can intersect line $DE$ at maximum 1 point  for $t\ge b$. Indeed, note that $\phi(b)=d-b+\frac{d(c-a)}{\sqrt{b^2-d^2}}$ and $\phi(+\infty)=-\infty$. Therefore, if $d-b+\frac{d(c-a)}{\sqrt{b^2-d^2}}\ge0$, then equation $\phi(t)=0$ has a unique solution in interval $t\ge b$. Otherwise, if $d-b+\frac{d(c-a)}{\sqrt{b^2-d^2}}<0$, then equation $\phi(t)=0$ has no solutions in interval $t\ge b$. In the statement of the problem we are given acute angled triangle $CDE$ satisfying all the conditions for extrema. Therefore, $d-b+\frac{d(c-a)}{\sqrt{b^2-d^2}}>0$ and the only solution of equation $\phi(t)=0$ in interval $t>b$ is $t_0=|AD|$. Note that $t_0^2=d^2+\lambda^2$, where $\lambda$ is one of the solutions of fourth degree equation
$$
\left(b^{2}-d^{2}\right) x^{4}+\left(2 b d a-2 b c\right) x^{3}+\left(-2 b^{3} d+2 b^{2} d^{2}+2 b \,d^{3}-2 d^{4}\right) x^{2}
$$
$$
+\left(-2 a \,b^{2} d^{2}+2 a b \,d^{3}+2 b^{2} c d-2 b c \,d^{2}\right) x
$$
$$
+a^{2} b^{2} d^{2}+b^{4} d^{2}-2 b^{3} d^{3}+2 b \,d^{5}-d^{6}-2 a \,b^{2} c d+b^{2} c^{2}=0,
$$
and therefore can be solved in radicals using Ferrari's method.

It remains to show that if $b>d$ and $k_2>0$, then
$$
\lim_{t\rightarrow 0^+}f(a+k_1t , b+k_2t)>|CD|+|CE|.
$$
Since the limit is minimal when $k_1=0$, it is sufficient to show that
$$
\lim_{y\rightarrow b^+}f(a ,y)>|CD|+|CE|.
$$
Note that since $2d^{3}+y^{3}\ge 3 d^{2} y$,
$$
f_{yy}(a,y)=\frac{\left(a-c\right)^{2} \left(2d^{3}+{y^{3}}-{3 d^{2} y}+2 d\left(a-c\right)^{2}\right)}{\sqrt{ \left(\left(y-d\right)^{2}+\left(a-c\right)^{2}\right)^3}\, \left(y-d\right)^{3}}>0,
$$
and therefore $f(a,y)$ is convex in interval $y\ge b$. Also $
\lim_{y\rightarrow +\infty}f_{y}(a,y)=2>0$ and
$$
\lim_{y\rightarrow b^+}f_{y}(a,y)=1+\sqrt{1+\frac{\left(a-c\right)^{2}}{\left(b-d\right)^{2}}}-\frac{b \left(a-c\right)^{2}}{\sqrt{1+\frac{\left(a-c\right)^{2}}{\left(b-d\right)^{2}}}\, \left(b-d\right)^{3}}<0.
$$
Indeed, 
$$
\lim_{y\rightarrow b^+}f_{y}(a,y)=\left(1+\sqrt{1+\frac{\left(a-c\right)^{2}}{\left(b-d\right)^{2}}}\right)\left(1-\frac{b}{b-d}\frac{\sqrt{1+\frac{\left(a-c\right)^{2}}{\left(b-d\right)^{2}}}-1}{\sqrt{1+\frac{\left(a-c\right)^{2}}{\left(b-d\right)^{2}}}}\right),
$$
where the first factor is always positive. The second factor can be simplified as
$$
\frac{b(b-d)-d\sqrt{\left(a-c\right)^{2}+\left(b-d\right)^{2}}}{\left(b-d\right)\sqrt{\left(a-c\right)^{2}+\left(b-d\right)^{2}}}
$$
$$
=\frac{(b^2-d^2)(b-d)^2-d^2(a-c)^2}{(b-d)\sqrt{\left(a-c\right)^{2}+\left(b-d\right)^{2}}\left(b(b-d)+d\sqrt{\left(a-c\right)^{2}+\left(b-d\right)^{2}}\right)},
$$
where the denominator is always positive. The numerator can be factorised as
$$
\left((b-d)\sqrt{b^2-d^2}-d(a-c)\right)\left((b-d)\sqrt{b^2-d^2}+d(a-c)\right),
$$
which is negative because as was mentioned before $d-b+\frac{d(c-a)}{\sqrt{b^2-d^2}}>0$.
Therefore $f(a,y)$ is decreasing and then increasing for $y>b$ and consequently, $f(a,y)$ can not have infimum at $(a,b)$ and the only minimum of $f(x,y)$ is the point $C$.
\end{proof} 
As an immediate application of Theorem 1 we obtain the following particular result.

\begin{cor} Let $CDE$ be a triangle with acute angles at the vertices $D$ and $E$, such that $|CE|>|CD|$. Let $CK$ be the angle bisector of the triangle $CDE$. Drop the perpendiculars $KL$ and $KM$ to the sides $CD$ and $CE$, respectively. Take the points $A$ and $B$ on the sides $CD$ and $CE$, respectively, such that $|AC|=|LD|$, $|BC|=|ME|$. Drop the perpendicular $AA_0$, $BB_0$, and $CC_0$ to side $DE$. Extend line $AB$ to intersect the line $DE$ at the point $P$. Then
$|AA_0|+|AP|>|CD|+|CE|$ (see Figure \ref{fig8}).
\end{cor}

Note that we also solved the following more general problem.

\begin{problem} Given \(A(a,b),\ B(c,d),\) with $a< c$ and $b\ge d>0$ find the minimum of $|C_1 D_1 |+|C_1 E_1|$, where \(C_{1}(x,y)\) is a point such that $y>b$ and the rays $C_1 A$ and $C_1 B$ intersect the line $y=0$ at the points $D_1$ and $E_1$, respectively.
\end{problem}

\begin{figure}[htbp]
\centerline{\includegraphics[scale=.3]{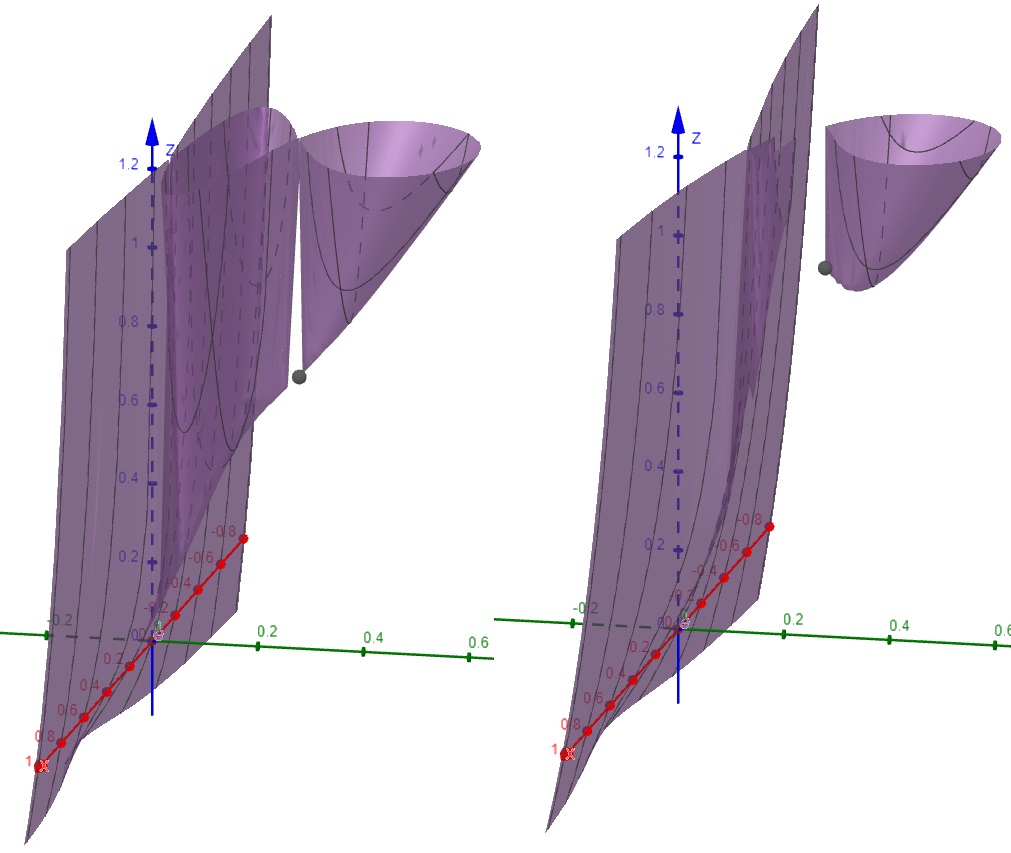}}
\caption{Graph of $f$ in cases $d-b+\frac{d(c-a)}{\sqrt{b^2-d^2}}\le0$ (left, $a=0.1$, $b=0.3$, $c=0.3$, $d=0.1$) and $d-b+\frac{d(c-a)}{\sqrt{b^2-d^2}}>0$ (right, $a=0.1$, $b=0.3$, $c=0.3$, $d=0.2$). Black point $\left(a,b,\lim_{y\rightarrow b^+}{f(a,y)}\right)$.}
\label{fig7}
\end{figure}

\begin{figure}[htbp]
\centerline{\includegraphics[scale=.15]{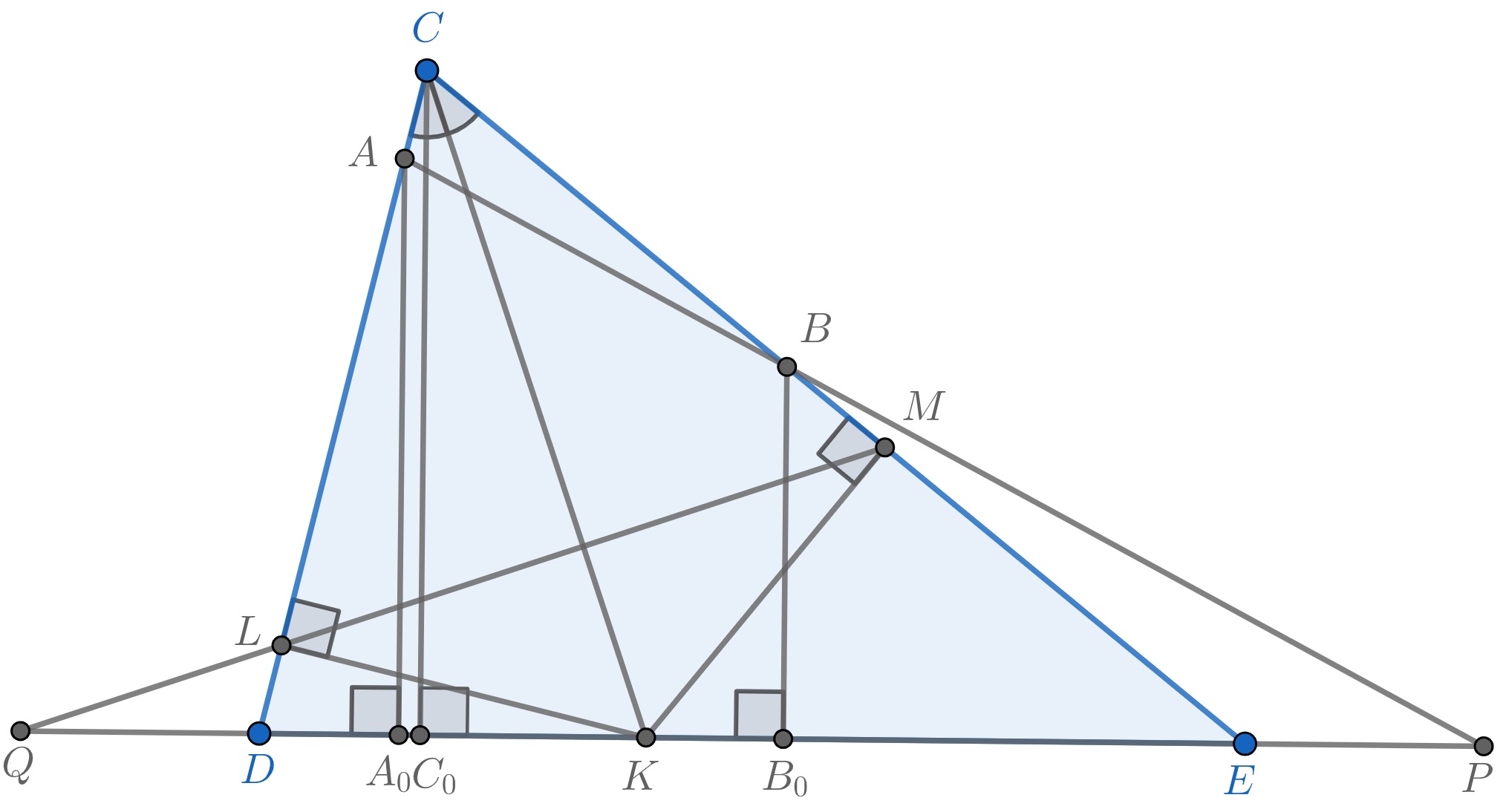}}
\caption{Properties of $\triangle CDE$.}
\label{fig8}
\end{figure}

\noindent\textbf{Answer.} If $b=d$ or if $b>d$ and $d-b+\frac{d(c-a)}{\sqrt{b^2-d^2}}>0$, then the minimum exists and it is equal to $|CD|+|CE|$, from the previous result. If $b>d$ and $d-b+\frac{d(c-a)}{\sqrt{b^2-d^2}}\le0$, then the minimum does not exist. In this case there is infimum of $|C_1 D_1 |+|C_1 E_1|$ at point $(a,b)$ which is equal to $b \left(1+\sqrt{1+\frac{\left(a-c\right)^{2}}{\left(b-d\right)^{2}}}\right)$. See Figure \ref{fig7} for the graph of $f(x,y)$ in these two cases.

\noindent\textit{Remark} 1. Note that in general \(f(x,y)\) is not a convex function (cf. \cite{zas}). So, we can not use the well know result in analysis which says that a local minimum of a convex function is also the absolute minimum (see e.g. \cite{ciar}, p. 543). Actually, each of the functions
$$
f_1\left(x,y\right)= y\cdot \sqrt{1+\frac{\left(x-a\right)^{2}}{\left(y-b\right)^{2}}},\ 
f_2\left(x,y\right)= y\cdot \sqrt{1+\frac{\left(x-c\right)^{2}}{\left(y-d\right)^{2}}},
$$
have negative Hessian. Indeed,
$$
(f_1)_{xx}(f_1)_{yy}-(f_1)_{xy}(f_1)_{yx}=-\frac{b^{2} \left(x-a\right)^{2}}{\left(y-b\right)^{4} \left(\left(x-a\right)^{2}+(y-b)^{2}\right)}<0,
$$
and similarly for $f_2$. There are points $(x,y)$ for which Hessian of $f=f_1+f_2$ is also negative. For example, for the points of the set $\mathbb{D}$ which are on the line $AB$ we can write $x= \left(\lambda +1\right) a-\lambda  c$ and $y= (\lambda +1) b -\lambda  d $, where $\lambda>0$, and obtain
$$
(f)_{xx}(f)_{yy}-(f)_{xy}(f)_{yx}=-\frac{\left(\left(\lambda +1\right)^{2} b+d \lambda^{2}\right)^{2} \left(a-c\right)^{2}}{\left(b-d\right)^{4} \lambda^{4} \left(\left(a-c\right)^{2}+\left(b-d\right)^{2}\right) \left(\lambda +1\right)^{4}}<0.
$$
Similarly, this Hessian is negative for the points of the set $\mathbb{D}$ which are close to the line $y=b$.

\noindent\textit{Remark} 2. We proved that the triangle $CDE$ has optimal $|CD|+|CE|$ within all the other triangles $C_1D_1E_1$, which share common pair of points $A$ and $B$ on their sides. The triangle $CDE$ has also other interesting properties worth mentioning. First drop the perpendiculars $AA_0$, $BB_0$, and $CC_0$ to the side $DE$ and suppose that $|CD|<|CE|$. Extend the lines $AB$ and $ML$ to intersect the line $DE$ at the points $P$ and $Q$, respectively (see Figure \ref{fig8}).
\begin{itemize}
\item $|PE|=|QD|=\frac{|DE|^2}{|B_0C_0|-|A_0C_0|}$. In particular, this implies that if the points $D$, $E$, and $C_0$ are fixed, and the point $C$ moves vertically along $CC_0$, then the lines $AB$ and $LM$ pass though the fixed points $P$ and $Q$, respectively.
\item $|DE|=|A_0B_0|+\frac{[CDE]}{R}$, where $[CDE]$ and $R$ are the area and the circumradius, respectively, of $\triangle CDE$.
\item $\frac{|AA_0|}{|BB_0|}=\frac{|CE|}{|CD|}$.
\item $|CA|+|CB|=\frac{|DE|^2}{|CE|+|CD|}$.
\item $|PQ|=\frac{|DE|^3}{|CE|^2-|CD|^2}$.
\item $\frac{|CA|}{|CB|}=\frac{|DC_0|}{|EC_0|}$.

\end{itemize}

\noindent\textit{Remark} 3. It is possible to generalize Problem 1 by asking to find the minimum of $\left(|C_1 D_1 |^p+|C_1 E_1|^p\right)^{\frac{1}{p}}$ $(p\ne 0)$. If $p\ge 1$, then this expression can be interpreted as $l_p$ norm (\cite{lust}, p. 20). If $p\rightarrow 0$, then $$\left(|C_1 D_1 |^p+|C_1 E_1|^p\right)^{\frac{1}{p}}\rightarrow \sqrt{|C_1 D_1 |\cdot|C_1 E_1|}.$$We already solved the case $p=1$. In analogy with \cite{prot1}, \cite{zas} we can look at the case $p=\infty$. In this case  $$\left(|C_1 D_1 |^p+|C_1 E_1|^p\right)^{\frac{1}{p}}=\max{\left(|C_1 D_1 |,|C_1 E_1|\right)}$$and the solution is essentially the same. We will only give the answer and some implications.
\begin{problem} Given \(A(a,b),\ B(c,d),\) with $a< c$ and $b\ge d>0$ find the minimum of $\max{\left(|C_1 D_1 |,|C_1 E_1|\right)}$, where \(C_{1}(x,y)\) is a point such that $y>b$ and the rays $C_1 A$ and $C_1 B$ intersect the line $y=0$ at the points $D_1$ and $E_1$, respectively.
\end{problem}

\begin{figure}[htbp]
\centerline{\includegraphics[scale=.2]{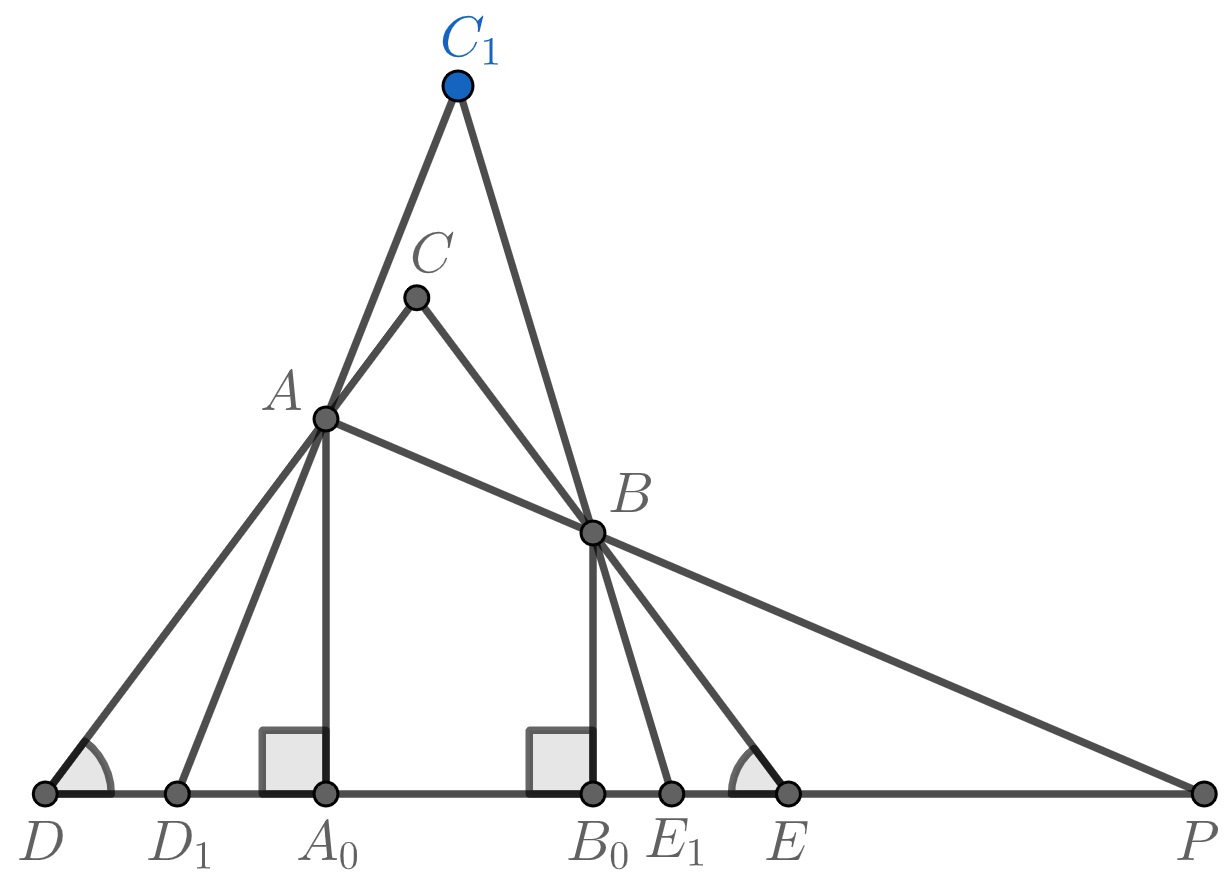}}
\caption{Inequalities $\max{\left(|C_1 D_1 |,|C_1 E_1 |\right)}>|CD|$ and $|AP|> |CD|$.}
\label{fig9}
\end{figure}

\noindent\textbf{Answer.}  If $k_0:=\left(\frac{b+d}{c-a}\right)^{\frac{1}{3}}>\frac{b-d}{c-a}$, then
the minimum of function $f(x,y)$ is $f(x^{*},y^{*})$, where
$$
x^{*}=\frac{a+c}{2}-\frac{b-d}{2 k_0},\ y^{*}=\frac{b+d}{2}-k_0\frac{a-c}{2}.
$$
If $k_0\le\frac{b-d}{c-a}$, then the minimum of function $f(x,y)$ does not exist. In this case there is infimum of function $f(x,y)$ at the points infinitely close to $(a,b)$ and the value of the infimum is equal to
$$
b\sqrt{\left(\frac{c-a}{b-d}\right)^2+1}.
$$
Using this result we can prove the following theorem.
\begin{theorem} Let $A$ and $B$ be points on the sides $CD$ and $CE$, respectively, of the isosceles triangle $CDE$ such that $\tan{\angle D}=\tan{\angle E}=\left(\frac{|AA_0|+|BB_0|}{|A_0B_0|}\right)^{\frac{1}{3}}$, where $AA_0$ and $BB_0$ are the perpendiculars from points $A$ and $B$, respectively, to the side $DE$, and $|AA_0|>|BB_0|$ (see Figure \ref{fig9}).  Then for any point $C_1$ different from $C$, such that the rays $C_1 A$ and $C_1 B$ intersect the line $DE$ at the points $D_1$ and $E_1$, respectively, the inequality $\max{\left(|C_1 D_1 |,|C_1 E_1 |\right)}>|CD|$ holds true. In particular, if the lines $AB$ and $DE$ intersect at the point $P$, then $|AP|>|CD|$ (cf. \cite{aliyev2}).
\end{theorem}

I leave as an open question to study the cases $p=0$, $p=2$ and the other values of $p$. It seems that $p=1$ is a unique case when the inequality is always true.
\begin{problem} (Open) Let $p\ne1$ be a real number. Let $CK$ be a cevian of the triangle $CDE$ such that $\frac{|DK|}{|KE|}=\left(\frac{|DC|}{|CE|}\right)^p$. Drop the perpendiculars $KL$ and $KM$ to the lines $CD$ and $CE$, respectively. Take the points $A$ and $B$ on the lines $CD$ and $CE$, respectively, such that $\overrightarrow{AC}=\overrightarrow{DL}$, $\overrightarrow{BC}=\overrightarrow{EM}$. Take any point $C_1$ different from $C$, such that the rays $C_1 A$ and $C_1 B$ intersect the line $DE$ at the points $D_1$ and $E_1$, respectively. Does the inequality $$\left(|C_1 D_1 |^p+|C_1 E_1|^p\right)^{\frac{1}{p}}>\left(|CD|^p+|CE|^p\right)^{\frac{1}{p}}$$hold true for all possible triangles $CDE$ and points $C_1$ (see Figure \ref{fig1})?
\end{problem}

In particular, for the case $p=2$ one can show that if $CK$ of Problem 3 is a symedian, then $$|C_1 D_1 |^2+|C_1 E_1 |^2>|CD|^2+|CE|^2$$ is not always true. One can find counterexamples by considering triangles $CDE$ with small $|CD|$ and the point $C_1$ close to the point $A$. The case $p<1$ seems to be similar because the point $C$ in this case is not always an absolute extremum point. The expression $\left(|C_1 D_1 |^p+|C_1 E_1|^p\right)^{\frac{1}{p}}$ can get smaller (if $0\le p<1$) and greater (if $p<0$) values when point $C_1$ approaches points $A$ or $B$. For example, in the case $p=0$, if $CK$ of Problem 3 is a median, then the inequality $|C_1 D_1 |\cdot |C_1 E_1 |>|CD|\cdot |CE|$ is not always true. As a problem for further exploration it would also be interesting to study these questions for more than two points instead of just the points $A$ and $B$. One can also replace the line $DE$ with a plane and ask the same question for more than two points in space.
\section{Conclusion}
In the paper we solved geometric optimization problems inspired by Philo's line. The obtained solution reveals interesting properties of the angle bisector and its projections on the sides of this triangle. The provided proof uses multivariable optimization and geometrical methods. Some generalizations and open problems, in particular, the impossibility of the analogous inequality for medians and symedians are discussed.

\section*{Appendix} In this part we will give an elementary solution for the classical problem mentioned in the introduction above. In the literature, this problem is usually solved using calculus. A different elementary proof based on angle chasing is given in \cite{anghel2}.
\begin{problem}
Let the points $D$ and $G$ on the side $EF$ of $\triangle BEF$ such that the points $D$ and $G$ are symmetric with respect to the center of the side $EF$ and $BG\perp EF$. Let $E_1F_1$ be any line through point $D$ and different from $EF$ such that the points $E_1$ and $F_1$ are on the rays $BE$ and $BF$, respectively. Prove that $|E_1F_1|>|EF|$.
\end{problem}
\begin{figure}[htbp]
\centerline{\includegraphics[scale=.2]{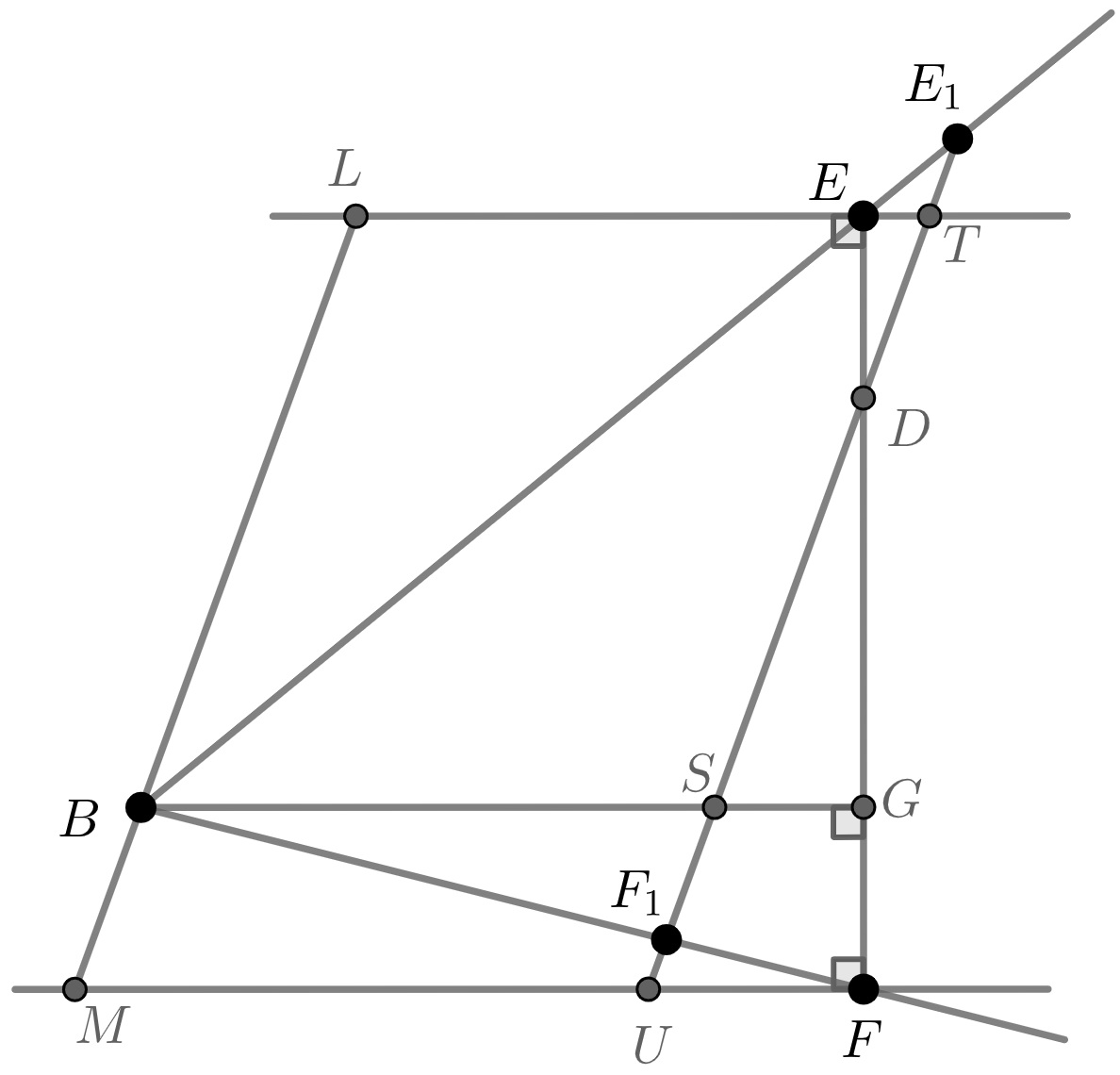}}
\caption{ Proof of the inequality $|E_1F_1|>|EF|$.}
\label{fig10}
\end{figure}

\noindent\textit{Solution.} Let the lines $BG$ and $E_1F_1$ meet at the point $S$ (see Figure \ref{fig10}). Let the lines through $E$ and $F$ parallel to $BG$ intersect the line $E_1F_1$ at the points $T$ and $U$, respectively. Let these parallel lines also intersect the line through $B$ and parallel to $E_1F_1$ at the points $L$ and $M$, respectively. Then $\frac{|LB|}{|BM|}=\frac{|EG|}{|GF|}=\frac{|DF|}{|DE|}$. By the similarity of $\triangle EE_1T$ and $EBL$, $\frac{|E_1T|}{|BL|}=\frac{|ET|}{|EL|}$. Analogously, by the similarity of $\triangle FF_1U$ and $FBM$, $\frac{|BM|}{|F_1U|}=\frac{|MF|}{|UF|}$. Using the similarity of $\triangle ETD$, $\triangle GSD$ and $\triangle FUD$, we find $|UF|=|ET|\cdot\frac{|DF|}{|DE|}$ and  $|SG|=|ET|\cdot\frac{|DG|}{|DE|}$. Since $MUSB$ and $SBLT$ are parallelograms, $
|MU|=|BS|=|LT|=|BG|-|ET|\cdot \frac{|DG|}{|DT|}$. So, $|MF|=|MU|+|UF|=|BG|+|ET|$ and $|LE|=|LT|-|ET|=|BG|-|ET|\cdot\frac{|EG|}{|DE|}$. By mutiplying the expressions for $\frac{|E_1T|}{|BL|}$, $\frac{|LB|}{|BM|}$, and $\frac{|BM|}{|F_1U|}$, we obtain $\frac{|E_1T|}{|F_1U|}=\frac{|BG|+|ET|}{|BG|-|ET|\cdot\frac{|EG|}{|DE|}}>1$. Therefore ${|E_1T|}>{|F_1U|}$ and consequently, ${|E_1F_1|}=|E_1T|+|TU|-|UF_1|>|E_1T|+|EF|-|UF_1|>|EF|.$ All other positions of the line $E_1F_1$ and the points $D$ and $G$ can be considered similarly.

\section*{Acknowledgments}
I thank Pyrkov V.E. for informing me about the existence of the paper \cite{mord}. 
I also thank Sinkevich G.I. for the opportunity to give a talk about the topic in The Seminar on the History of Mathematics \cite{aliyev3}.
\section{Declarations}
\textbf{Ethical Approval.}
Not applicable.
 \newline \textbf{Competing interests.}
None.
  \newline \textbf{Authors' contributions.} 
Not applicable.
  \newline \textbf{Funding.}
This work was completed with the support of ADA University Faculty Research and Development Fund.
  \newline \textbf{Availability of data and materials.}
Not applicable
% ------------------------------------------------------------------------

% ------------------------------------------------------------------------
\end{document}